\documentclass[preprint,11pt,nofootinbib]{revtex4-1}

\usepackage{graphicx}
\usepackage{dcolumn} 
\usepackage{bm}
\usepackage{bbm}
\usepackage{amssymb}
\usepackage{amsmath}
\usepackage{epsfig}    
\usepackage{color}
\usepackage{slashed}
\usepackage{hyperref}

\begin{document} 

\title{Explanation of the $W$ mass shift at CDF II in the extended Georgi-Machacek model}
\preprint{OU-HET-1143}
\date{\today}

\author{Ting-Kuo Chen}
\email{d10222031@ntu.edu.tw}
\affiliation{Department of Physics and Center for Theoretical Physics, National Taiwan University, Taipei, Taiwan 10617, Republic of China}

\author{Cheng-Wei Chiang}
\email{chengwei@phys.ntu.edu.tw}
\affiliation{Department of Physics and Center for Theoretical Physics, National Taiwan University, Taipei, Taiwan 10617, Republic of China}
\affiliation{Physics Division, National Center for Theoretical Sciences, Taipei, Taiwan 10617, Republic of China}

\author{Kei Yagyu}
\email{yagyu@het.phys.sci.osaka-u.ac.jp}
\affiliation{Department of Physics, Osaka University, Toyonaka, Osaka 560-0043, Japan}

\begin{abstract}

The CDF II experiment has recently determined the mass of the $W$ boson to be $m_W(\text{CDF II}) = 80.4335 \pm 0.0094$~GeV, which deviates from the standard model prediction at $7\sigma$ level.  Although this new result is in tension with other experiments such as those at LHC and LEP, it is worth discussing possible implications on new physics by this anomaly.  We show that this large discrepancy can be explained by nonaligned vacuum expectation values of isospin triplet scalar fields in the Georgi-Machacek model extended with custodial symmetry-breaking terms in the potential.  The latter is required to avoid an undesirable Nambu-Goldstone boson as well as to be the consistent treatment of radiative corrections.  With $m_W(\text{CDF II})$ as one of the renormalization inputs at the one-loop level, we derive the required difference in the triplet vacuum expectation values, followed by a discussion of phenomenological consequences in the scenario.

\end{abstract}
\maketitle

\section{Introduction \label{sec:Introduction}}

A new measurement of the $W$ boson mass has been recently reported by the CDF II Collaboration with an integrated luminosity of 8.8~fb$^{-1}$ at Tevatron.  
Surprisingly the measured value, $m_W(\text{CDF II}) = 80.4335 \pm 0.0094$~GeV
deviates significantly from the prediction of the standard model (SM), $m_W(\text{SM}) = 80.357\pm 0.006$~GeV, at $7\sigma$ level~\cite{CDF:2022hxs}. 
What is curious is that all the previous experiments at LHC and LEP and even the D0 II Collaboration at Tevatron 
show consistent results with the SM prediction, while the new CDF II result poses a tension with these measurements. 
Although such tension may originate from the systematics of experiments, it is worth exploring possible new physics explanations for the CDF II result and studying the implications thereof.

In this work, we show that the anomaly in the $W$ boson mass can be explained in models which break the custodial symmetry by nonaligned vacuum expectation values (VEVs) of isospin triplet Higgs fields.   
So far, several papers in this context have already existed in the literature~\cite{Cheng:2022jyi,Du:2022brr,Mondal:2022xdy,FileviezPerez:2022lxp,Kanemura:2022ahw,Popov:2022ldh,Batra:2022org,Heeck:2022fvl}.   In particular, we restrict ourselves to the Georgi-Machacek (GM) model~\cite{Georgi:1985nv,Chanowitz:1985ug} extended by introducing custodial symmetry-breaking terms in the potential.  The latter must be introduced to ensure a theoretically consistent framework, 
but is overlooked by some related papers that aim to address the CDF II anomaly~\cite{Du:2022brr,Mondal:2022xdy}~\footnote{Ref.~\cite{Du:2022brr} simply takes that the two triplets have misaligned vacuum expectation values at tree level, which is a theoretically inconsistent assumption as argued below.  In Ref.~\cite{Mondal:2022xdy}, 
the alignment of the triplet VEVs is imposed such that the electroweak $\rho$ parameter takes unity at tree level, 
but there is no dedicated discussion for its loop corrections. 
}.

This paper is organized as follows.  Sec.~\ref{sec:The Geoergi-Machacek Model} reviews the GM model and motivates why one needs to go beyond the model in order to accommodate the $W$ boson mass anomaly.  In Sec.~\ref{sec:Extended Georgi-Machacek Model}, we propose a minimal extension to the GM model that provides a self-consistent framework for the analysis.  Sec.~\ref{sec:Discussions} discusses the most prominent implications of the extended model.  The findings are summarized in Sec.~\ref{sec:Conclusions}.

\section{The Georgi-Machacek Model \label{sec:The Geoergi-Machacek Model}}

The scalar sector of the GM model is composed of an isospin doublet $\phi$ with hypercharge $Y = 1/2$, a complex triplet $\chi$ with $Y=1$ and a real triplet $\xi$ with $Y=0$.  The scalar potential is constructed to possess the custodial symmetry, i.e., to be invariant under a global $SU(2)_L\times SU(2)_R$ symmetry.  With the introduction of bidoublet and bitriplet fields
\begin{align}
\Phi = (\phi^c,\phi),\quad 
\Delta = (\chi^c,\xi, \chi)
, 
\end{align}
where $\phi^c$ and $\chi^c$ denote the charge-conjugated fields, the most general scalar potential that respects all the required symmetries is
\begin{align}
V_{\text{cust}}
=&
m_\Phi^2\text{tr}(\Phi^\dagger\Phi)+m_\Delta^2\text{tr}(\Delta^\dagger\Delta)
+\lambda_1[\text{tr}(\Phi^\dagger\Phi)]^2+\lambda_2[\text{tr}(\Delta^\dagger\Delta)]^2
+\lambda_3\text{tr}[(\Delta^\dagger\Delta)^2]\notag\\
&+\lambda_4\text{tr}(\Phi^\dagger\Phi)\text{tr}(\Delta^\dagger\Delta)
+\lambda_5\text{tr}\left(\Phi^\dagger\frac{\tau^a}{2}\Phi\frac{\tau^b}{2}\right)
\text{tr}(\Delta^\dagger t^a\Delta t^b)\notag\\
&+\mu_1\text{tr}\left(\Phi^\dagger \frac{\tau^a}{2}\Phi\frac{\tau^b}{2}\right)(P^\dagger \Delta P)^{ab}
+\mu_2\text{tr}\left(\Delta^\dagger t^a\Delta t^b\right)(P^\dagger \Delta P)^{ab}
~, \label{eq:pot}
\end{align}
where $\tau^a/2$ and $t^a$ ($a=1,2,3$) are the fundamental and triplet representations of the $SU(2)$ generators, respectively, and the matrix $P$ is given by 
\begin{align}
P=\left(
\begin{array}{ccc}
-1/\sqrt{2} & i/\sqrt{2} & 0 \\
0 & 0 & 1 \\
1/\sqrt{2} & i/\sqrt{2} & 0
\end{array}\right)
~. 
\end{align}
When the VEVs of doublet and triplet fields are taken to be in diagonal forms, i.e., 
$\langle\Phi\rangle = v_\Phi \mathbbm{1}_{2\times 2}$ and $\langle\Delta\rangle = v_\Delta \mathbbm{1}_{3\times 3}$, 
the $SU(2)_L\times SU(2)_R$ symmetry is spontaneously broken down to the diagonal one $SU(2)_V$, preserving 
the custodial symmetry.  In this case, the electroweak (EW) $\rho$ parameter at tree level $\rho_{\rm tree}$ is unity as a consequence of the $SU(2)_V$ symmetry.  The scalars can be classified under the $SU(2)_V$ symmetry into an $h$, representing the 125-GeV Higgs boson, a singlet $H_1^0$, a triplet $H_3^{0,\pm}$, and a quintet $H_5^{0,\pm,\pm\pm}$.

\begin{figure}[htbp]
\centering
\includegraphics[width=0.7\textwidth]{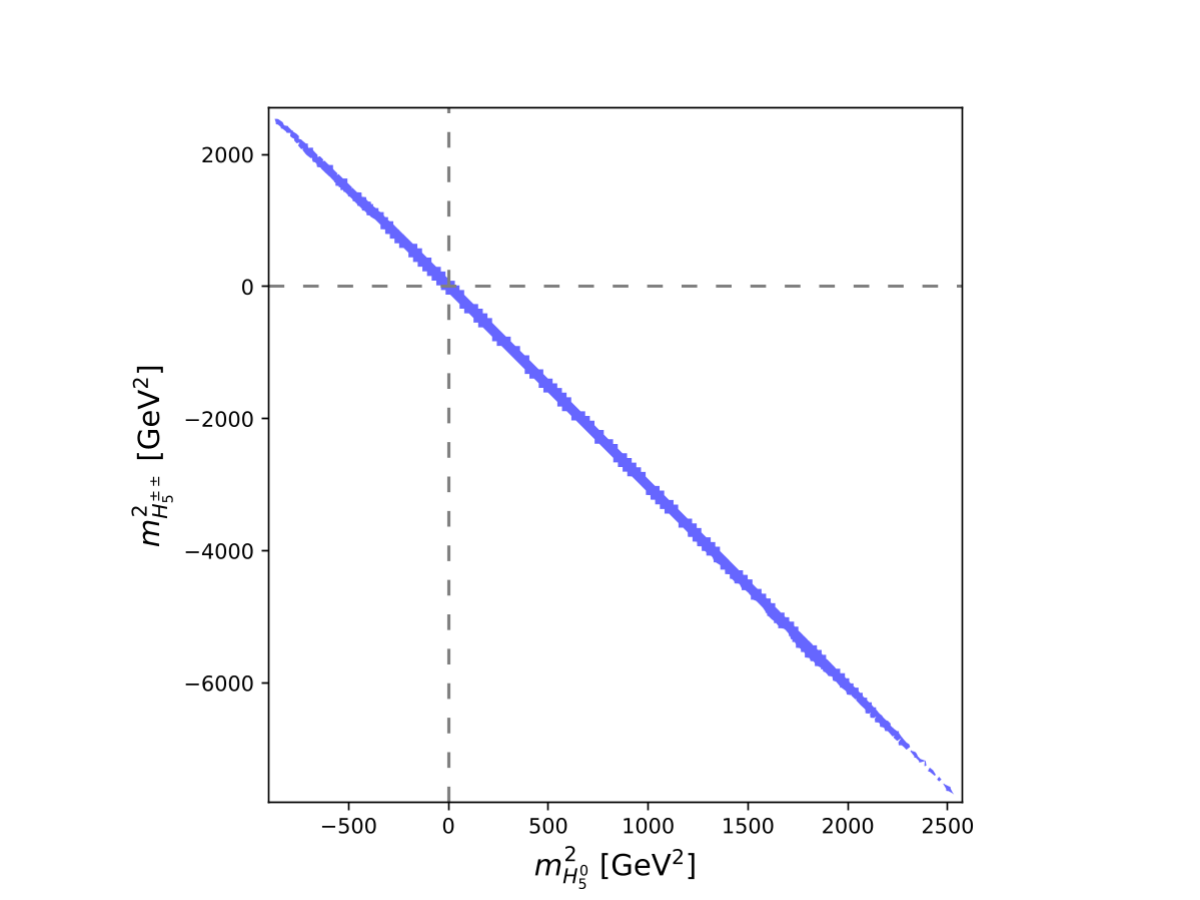}
\caption{\label{mHppmmsq_vs_mH1sq} Allowed region in the $m^2_{H_5^{\pm\pm}}$-$m^2_{H_5^0}$ plane, obtained by scanning the GM model parameters with misaligned triplet VEVs subject to the constraints of unitarity and perturbativity.}
\end{figure}

First, we would like to point out that at the tree level, the only sensible choice for the triplet fields is that they have an aligned VEV, i.e., $v_\xi \equiv \langle \xi^0\rangle$ and $v_\chi \equiv \langle \chi^0\rangle$ are the same and can be set as $v_\Delta$ defined above.  Had one chosen different VEVs $v_\chi \neq v_\xi$, the $SU(2)_V$ symmetry would be spontaneously broken down to a $U(1)$ symmetry which corresponds to the 
overall phase transformation of the $\Delta$ field, leading to the existence of two phenomenologically undesirable Nambu-Goldstone modes in the theory in addition to the usual ones absorbed into the longitudinal components of the $W^\pm$ and $Z$ bosons.  In fact, they would be the $H_5^\pm$ fields.  What is even worse is that a significant parameter space gives the mass relation $m^2_{H_5^{\pm\pm}} \simeq -3 m^2_{H_5^0}$ to a good approximation (when $|v_\chi - v_\xi| \ll 246$~GeV), as illustrated in Fig.~\ref{mHppmmsq_vs_mH1sq} that shows the allowed masses of $H_5^0$ and $H_5^{\pm\pm}$, under the constraints of perturbative unitarity~\cite{Aoki:2007ah,Hartling:2014zca} 
and vacuum stability~\cite{Hartling:2014zca} of the model. This is an indication that the theory is expanded around an unstable saddle point.  Besides, most of the parameter space even has the problem of breaking the $U(1)_{\rm em}$ symmetry because $m^2_{H_5^{\pm\pm}} < 0$.

In fact, it has been known for a long while that the custodial symmetry in the GM potential would be broken at loop level due to the hypercharge gauge boson and/or fermion loops~\cite{Gunion:1990dt,Blasi:2017xmc,Keeshan:2018ypw}. 
Thus, for a consistent renormalization prescription of the Higgs potential, one has to add terms that explicitly break the custodial symmetry in the potential from the very beginning~\cite{Chiang:2018xpl}. 
Therefore, in order to consistently discuss the GM model at the quantum level, at least one $SU(2)_L\times SU(2)_R$-breaking term must be added to $V_{\rm cust}$.  Also, only under this framework can one consider nonaligned triplet VEVs.

\section{Extended Georgi-Machacek Model \label{sec:Extended Georgi-Machacek Model}}

In the following, we consider a minimal extension of the Higgs potential by the replacement $m_\Delta^2 \text{tr}(\Delta^\dagger \Delta) \to m_\xi^2 \xi^\dagger \xi / 2 + m_\chi^2 \chi^\dagger \chi$.  That is, the scalar potential is now
\begin{align}
V = V_{\text{cust}}|_{m_\Delta \to 0} + \frac{m_\xi^2}{2} \xi^\dagger \xi + m_\chi^2 \chi^\dagger \chi
~, 
\end{align}
where we assume, in general, $v_\xi\neq v_\chi$.  With these triplet VEVs, the masses of weak bosons and the EW $\rho$ parameter are given at tree level as 
\begin{align}
m_W^2 &= \frac{g^2}{4}v^2,\quad m_Z^2 = \frac{g^2}{4c_W^2}(v^2 - \nu^2), \quad \rho_{\rm tree} = \frac{m_W^2}{m_Z^2c_W^2} = \frac{v^2}{v^2-\nu^2}
~, \label{eq:tree}
\end{align}
where $v^2 \equiv v_\Phi^2 + 4v_\chi^2 + 4v_\xi^2 = (\sqrt{2}G_F)^{-1}$, $\nu^2 \equiv 4(v_\xi^2 - v_\chi^2)$ and, $c_W^{}~(s_W^{})$ is the cosine (sine) of the weak mixing angle. 
Because of the additional parameter $\nu^2$ in the EW sector, we are allowed to select four EW parameters $\alpha_{\rm em}$, $G_F$, $m_Z^{}$ and $m_W^{}$ as inputs, 
with the first three being usually chosen as the EW input parameters in the SM.

At the tree level, the value of $\rho_{\rm tree}$ or, equivalently, that of $\nu^2$ is simply determined by Eq.~(\ref{eq:tree}), but its treatment can be different at loop levels. 
In order to discuss this issue, let us introduce $\Delta r$ which parametrizes a shift of the Fermi constant $G_F$ as $G_F \to G_F(1 - \Delta r)$ due to EW radiative corrections. 
The $\Delta r$ parameter consists of the following three parts~\cite{Bohm:1986rj}
\begin{align}
\Delta r =  \Delta \alpha_{\rm em} - \frac{c_W^2}{s_W^2}\Delta \rho  + \Delta r_{\rm rem}
~,  \label{eq:delr}
\end{align}
where $\Delta \alpha_{\rm em}$, $\Delta \rho$ and $\Delta r_{\rm rem}$ represent, respectively, radiative corrections to the fine structure constant, the $\rho$ parameter and the remaining part.  The individual parts can be explicitly expressed as 
\begin{align}
\Delta\alpha_{\rm em} &= \text{Re}\left[\frac{\Pi_{\gamma\gamma}'(0)}{m_Z^2} - \frac{\Pi_{\gamma\gamma}'(m_Z^2)}{m_Z^2}\right]
, \label{eq:delalpha} \\ 
\Delta\rho &= \text{Re}\left[\frac{\Pi_{ZZ}(0)}{m_Z^2} - \frac{\Pi_{WW}(0)}{m_W^2} + \frac{2s_W}{c_W}\frac{\Pi_{Z\gamma}(0)}{m_Z^2}\right] + \frac{\delta \rho}{\rho_{\rm tree}}
,\label{eq:delr4} \\ 
\Delta r_{\rm rem} & = \text{Re}\left[\frac{s_W^2 - c_W^2}{s_W^2}\frac{\Pi_{WW}(0)-\Pi_{WW}(m_W^2)}{m_W^2}+ \frac{c_W^2}{s_W^2}\frac{\Pi_{ZZ}(0)-\Pi_{ZZ}(m_Z^2)}{m_Z^2}  +\Pi_{\gamma\gamma}'(m_Z^2)\right] + \delta_{VB}
, \label{eq:delrem}
\end{align}
where $\Pi_{VV'}$ are one-particle irreducible (1PI) diagrams contributing to the transverse components of the gauge boson two-point functions and $\delta_{VB}$ denotes the vertex and box corrections to the light fermion scattering process. 
In Eq.~(\ref{eq:delr4}), $\delta \rho$ denotes the counterterm for the $\rho$ parameter, which does not appear in models with $\rho_{\rm tree} = 1$\footnote{Here, we mean models 
which automatically satisfy $\rho_{\rm tree}=1$ without taking any tunings or alignments. Thus, the GM model with $\rho_{\rm tree}=1$ does not belong to this class of models. }, e.g., the SM and two-Higgs-doublet models. 
We will discuss the renormalization condition to determine the $\delta \rho$ parameter in the next paragraph. 
We note that new physics contributions to $\Delta r$ can also be well described by the $S$, $T$ and $U$ parameters introduced in Ref.~\cite{Peskin:1991sw}, if the masses of new particles are much larger than $m_Z$ and the new particles feebly couple to SM light fermions, which is indeed the case in the GM model.  In such a model, new contributions to the $\Delta r$ parameter are written in terms of $S$, $T$, and $U$ as 
\begin{align}
\begin{split}
\Delta r_{\rm rem}^{\rm NP} & = \frac{g^2}{16\pi}\left(2\Delta S - \frac{c_W^2-s_W^2}{s_W^2}\Delta U \right) \simeq \frac{g^2}{8\pi}\Delta S
,
\\ 
\Delta \rho^{\rm NP} &= \alpha_{\rm em} \Delta T
, 
\end{split}
\end{align}
with $\Delta X \equiv X_{\rm NP} - X_{\rm SM} $ $(X = S,T,U)$ and $X_{\rm NP}$ and $X_{\rm SM}$ being predictions in the new physics model and the SM, respectively. 
We note that contributions to $\Delta \alpha_{\rm em}$ and $\Delta U$ are suppressed by the factor of $v^2/M_{\rm NP}^2$~\cite{Grinstein:1991cd} with respect to $\Delta S$, where $M_{\rm NP}$ denotes the typical mass scale of new physics.
In Ref.~\cite{Strumia:2022qkt}, a global fit analysis has been done by including the new CDF result, and it has been shown that typically we need $\Delta \rho \simeq 10^{-3}$ and/or $\Delta S \simeq -0.1$
in order to be within the 90\% confidence level region of the $\Delta\chi^2$ analysis.

The determination of $\delta \rho$ is as follows. 
The value of $m_W^{}$ with EW radiative corrections is expressed as 
\begin{align}
(m_W^2)_{\rm ren } = \frac{m_Z^2}{2}\rho_{\rm tree}\left[1 + \sqrt{1 - \frac{4\pi \alpha_{\rm em}}{\sqrt{2}G_Fm_Z^2\rho_{\rm tree}(1-\Delta r)}} \right]
~. \label{eq:mwren}
\end{align}
Since $\Delta r$ depends on $\delta \rho$ via the contribution of $\Delta \rho$,
one can impose the renormalization condition for determining the counterterm $\delta \rho$ as~\cite{Chiang:2018xpl} 
\begin{align}
\Delta \rho =  \Delta\rho_{\rm exp}
~, \label{eq:delrho}
\end{align}
such that 
\begin{align}
(m_W^2)_{\rm ren } = (m_W^2)_{\rm exp}
~. 
\end{align}
This condition can also be understood in such a way that the value of $\rho_{\rm tree}$ or, equivalently, $\nu^2$ is determined by fixing the $\Delta r$ parameter such that the observed value of $m_W$ is reproduced.

\begin{figure}[htbp]
\centering
\includegraphics[width=0.7\textwidth]{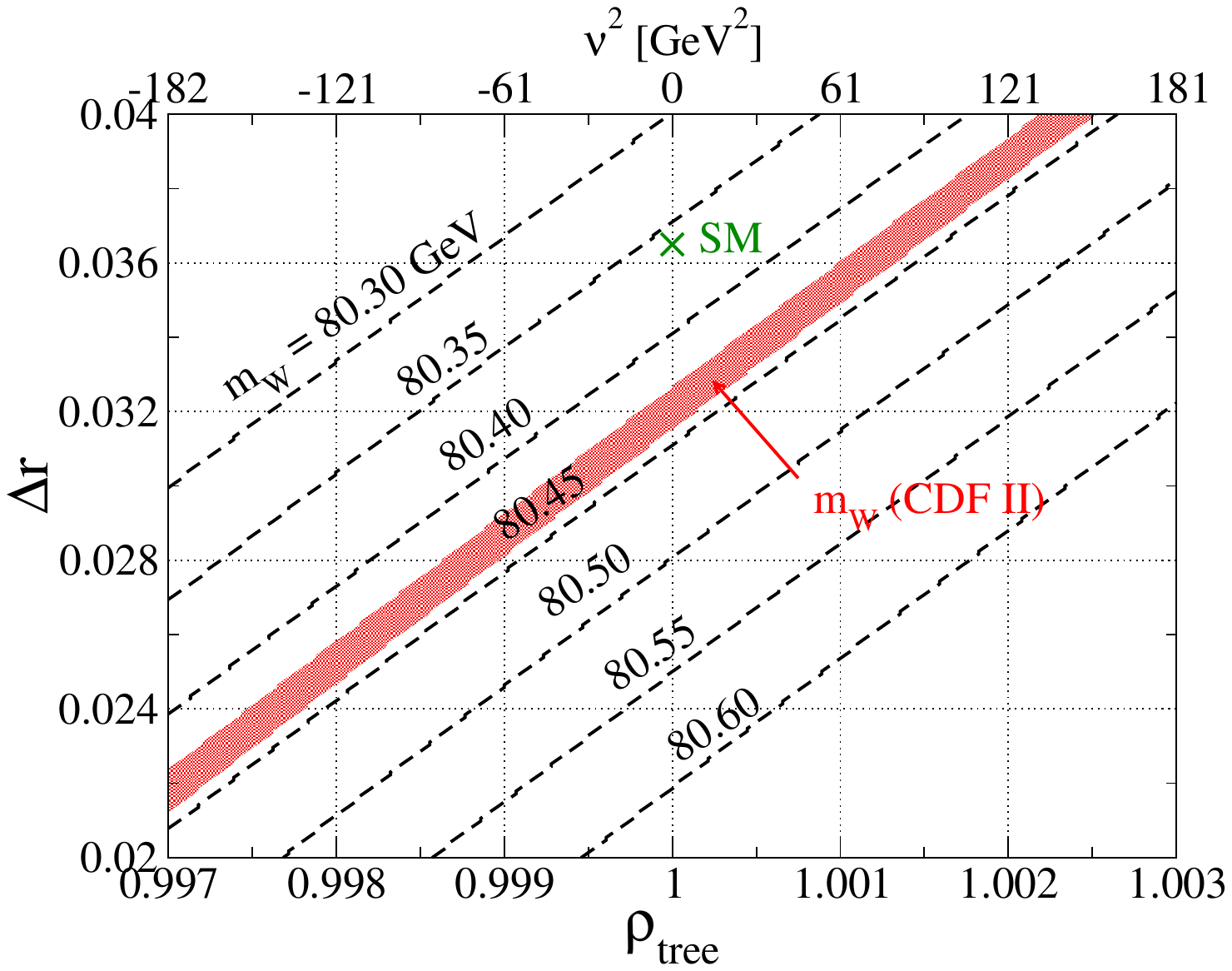}
\caption{Contour plots for the value of $m_W$ by using Eq.~(\ref{eq:mwren}) in the $\rho_{\rm tree}$--$\Delta r$ plane. The region shaded in red indicates the measured value at the CDF II experiment, i.e., 
$m_W(\text{CDF II}) = 80.4335 \pm 0.0094$~GeV. 
We also show the corresponding value of $\nu^2$ in the upper horizontal axis. }
\label{fig:mw}
\end{figure}

In Fig.~\ref{fig:mw}, we show the value of $m_W^{}$ as a function of $\rho_{\rm tree}$ and $\Delta r$ by using Eq.~(\ref{eq:mwren}), where the SM prediction is shown by the green cross, i.e., 
$\Delta r(\rm SM) = 0.0367141$ and $\rho_{\rm tree} = 1$ giving $m_W(\rm SM) = 80.357$~GeV~\cite{ParticleDataGroup:2020ssz}. 
In this plot, we use the following EW input values~\cite{ParticleDataGroup:2020ssz} and the $W$ boson mass measured at CDF II~\cite{CDF:2022hxs}, indicated by the red contour:
\begin{align}
\alpha_{\rm em}^{-1} = 137.036,~
G_F = 1.1663787\times 10^{-5}~\text{GeV}^{-2},~
m_Z=91.1876~\text{GeV},~
m_W = 80.4335~\text{GeV}
. 
\end{align} 
The plot shows that in order to accommodate the value of $m_W$ measured at CDF II, we need a negative contribution to $\Delta r$ and/or a positive shift of $\rho_{\rm tree}$ away from unity, where  
the former corresponds to a positive contribution from the $\Delta \rho$ parameter [see Eq.~(\ref{eq:delr})]. 
The direction of the shift from the SM prediction depends on the new physics model. 
For example, in two-Higgs-doublet models the prediction is shifted straight down. As shown in e.g., Refs.~\cite{Bahl:2022xzi, Lee:2022gyf}, 
the CDF II anomaly can be explained by taking the mass difference between a heavy neutral and a charged Higgs boson to be of the order of 100 GeV. 
In a model extended with only a $Y=0$ ($Y=1$) triplet Higgs field, the shift is toward the lower right (lower left), because such a model gives a positive (negative) shift in $\rho_{\rm tree}$. 
For instance, it has been shown in Ref.~\cite{FileviezPerez:2022lxp} that the CDF II anomaly can be explained by only the effect of $\rho_{\rm tree}$ by taking the real triple VEV to be about 5 GeV, 
which is consistent with our work. 
On the other hand, in a model extended with the $Y=1$ triplet field, a negative new contribution to $\Delta r$ is required to compensate for the effect of $\rho_{\rm tree}(\leq 1)$. One can achieve this by taking a mass splitting among the triplet-like Higgs bosons to be of the order of 100 GeV~\cite{Kanemura:2022ahw}. 
In the extended GM model, we can take either a positive or a negative shift of $\rho_{\rm tree}$, because the sign of $\nu^2$ is free. 
For example, if we take $\Delta r = \Delta r({\rm SM})$, we obtain $\nu^2=85.38\pm 10.09$~GeV$^2$, which corresponds to the case where the prediction is shifted from the SM value straight to the right.  

\section{Phenomenological implications \label{sec:Discussions}}

Let us discuss phenomenological implications of the parameter space favored by the CDF II anomaly. 
In particular, we focus on the deviation in the Higgs boson couplings to weak bosons from the SM predictions.  Denoting $g_{hVV}$ as the $hVV$ ($V=W,Z$) couplings in the model, the corresponding scale factors are defined as $\kappa_V\equiv g_{hVV}/g_{hVV}^{\rm SM}$, where $g_{hVV}^{\rm SM}$ are the SM couplings.  In our scenario,
\begin{align}
\kappa_W &= \frac{v_\Phi}{v}R_{11} + 4\frac{v_\xi}{v}R_{21} + 2\sqrt{2}\frac{v_\chi}{v}R_{31}
~, 
\label{kW}
\\
\kappa_Z &= \frac{v_\Phi}{v}R_{11} + 4\sqrt{2}\frac{v_\chi}{v}R_{31}
~, 
\label{kZ}
\end{align}
where $R_{ij}$ are the elements of the $3\times 3$ orthogonal matrix that connects the weak and mass eigenbases of the $CP$-even Higgs bosons: 
\begin{align}
\begin{pmatrix}
h_\phi \\
h_\xi\\
h_\chi
\end{pmatrix} = R 
\begin{pmatrix}
H_1 \\
H_2 \\
H_3
\end{pmatrix}
, 
\end{align}
with $h_\phi \equiv \sqrt{2}\text{Re}(\phi^0)-v_\Phi$, $h_\xi \equiv \xi^0-v_\xi$ and $h_\chi \equiv \sqrt{2}\text{Re}(\chi^0-v_\chi)$. 
We can identify the $H_1$ state as the discovered 125-GeV Higgs boson $h$. 
We note that in the limit where the custodial symmetry is restored in the tree-level potential, i.e., $v_\chi = v_\xi = v_\Delta$ and $V \to V_{\rm cust}$ given in Eq.~(\ref{eq:pot}), 
the mixing matrix $R$ reduces to~\cite{Chiang:2013rua} 
\begin{align}
R \xrightarrow[SU(2)_V]{} 
\begin{pmatrix}
1 & 0 &0 \\
0 & \frac{1}{\sqrt{3}} & -\sqrt{\frac{2}{3}}\\
0 & \sqrt{\frac{2}{3}} & \frac{1}{\sqrt{3}}
\end{pmatrix}
\begin{pmatrix}
 \cos\alpha & -\sin\alpha &0\\
 \sin\alpha &  \cos\alpha &0\\
0 & 0 & 1 
\end{pmatrix}
, 
\end{align}
as expected, and then both $\kappa_W$ and $\kappa_Z$ have the same value~\cite{Chiang:2013rua}
\begin{align}
\kappa_V &= \cos\beta \cos\alpha + \frac{2\sqrt{6}}{3}\sin\beta \sin\alpha
~, 
\end{align}
with $\tan\beta \equiv 2\sqrt{2}v_\Delta/v_\Phi$. 
The mixing angle $\alpha$ is determined by the potential parameters given in Eq.~(\ref{eq:pot}); see e.g., Ref.~\cite{Chiang:2013rua}.

In the present scenario without the custodial symmetry in the potential, we perform a parameter scan using the Bayesian-based global fitting package HEPfit~\cite{DeBlas:2019ehy}. We assign the priors of the parameters to be
\begin{equation}
	v_{\chi,\xi}\in[0,50]~{\rm GeV}, ~\lambda_{2-5}\in[-5,5], ~\mu_{1,2}\in[-3000,3000]~{\rm GeV} ,
\end{equation}
under the constraints of perturbative unitarity, vacuum stability, and uniqueness of the global minimum. 
Moreover, we take the CDF II measurement uncertainty into consideration, which translates to $\nu^2=85.38\pm 10.09$~GeV$^2$, or, equivalently, $v_\xi^2 - v_\chi^2 = 21.34\pm 2.52$~GeV$^2$. 
We note that $v_\Phi$ is determined by the relation $v^2 = v_\Phi^2 + 4v_\chi^2 + 4v_\xi^2$, and $\lambda_1$ is fixed to satisfy $m_{h}=125$ GeV for each scanned point. 
We further split the scanning range of $v_\chi$ into three intervals: $[0,0.5]$, $[0.5,5]$, and $[5,50]$~GeV, 
in order to see the $v_\chi$ dependence of $\kappa_V^{}$ and $\kappa_F$. 
In particular, the $v_\chi \to 0$ limit provides a similar scenario to the minimal Higgs triplet model composed of $\phi$ and $\xi$, 
which can also explain the CDF II anomaly only by the effect of the triplet VEV as mentioned in the previous section. 
Finally, in order to focus on the mass range that can be probed at the LHC, we impose an auxiliary constraint on the exotic scalar masses such that they are all below 2~TeV. We also assume that they are all heavier than 125~GeV to avoid the additional scalar decay modes for $H_1$ and, without loss of generality, that $m_{H_3}\geq m_{H_2}$.
We can then compare the predictions between our model and the minimal triplet model. 

\begin{figure}[htbp]
\centering
\includegraphics[width=0.7\textwidth]{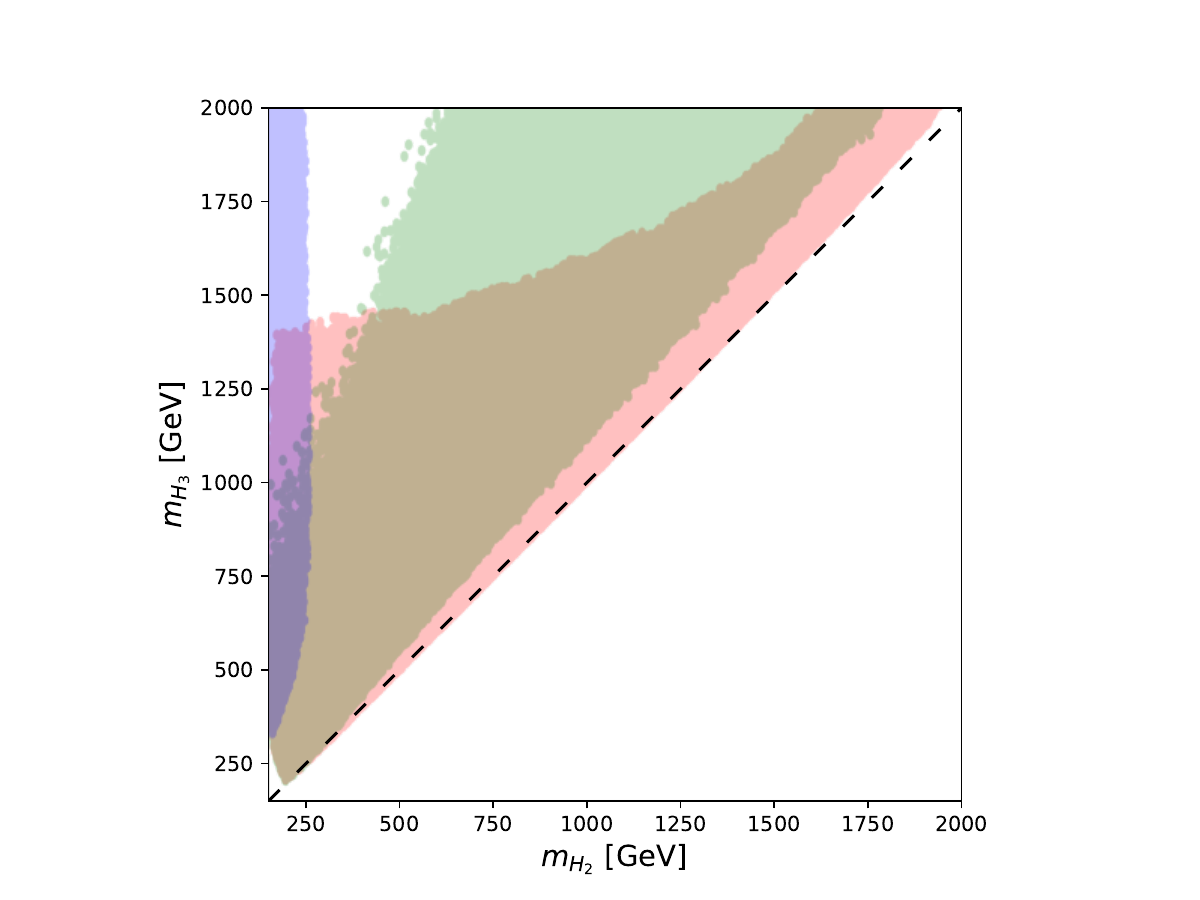}
\caption{Predicted values by the extended GM model with misaligned triplet mass terms in the $m_{H_2}$-$m_{H_3}$ plane.  The red, green and blue points show the case with $v_\chi\in[5,50]$~GeV, $v_\chi\in[0.5,5]$~GeV and $v_\chi\in[0,0.5]$~GeV, respectively.
}
\label{mH2_vs_mH3} 
\end{figure}

We present in Fig.~\ref{mH2_vs_mH3} the predicted values of $m_{H_2}$ and $m_{H_3}$ from the scan, 
where the red, green and blue points are allowed for the case with $v_\chi\in[5,50]$, [0.5,5, and [0,0.5]~GeV, respectively.  The corresponding $m_\chi^2$ and $m_\xi^2$ parameters span the range of $[-0.25,4] \times 10^6$~GeV$^2$ with $m_\xi^2 \le m_\chi^2$.  
We particularly point out the interesting feature that the blue points present a bound of $m_{H_2}\lesssim 250$~GeV. 
This behavior can be understood from Eq.~(\ref{eq:cp-even}), where the mass of the $\xi$-related field remains at the EW scale in the limit of $v_\chi \to 0$ with $v_\xi \neq 0$.  
We will discuss the decoupling limit in more detail later.

\begin{figure}[htbp]
\centering
\includegraphics[width=0.7\textwidth]{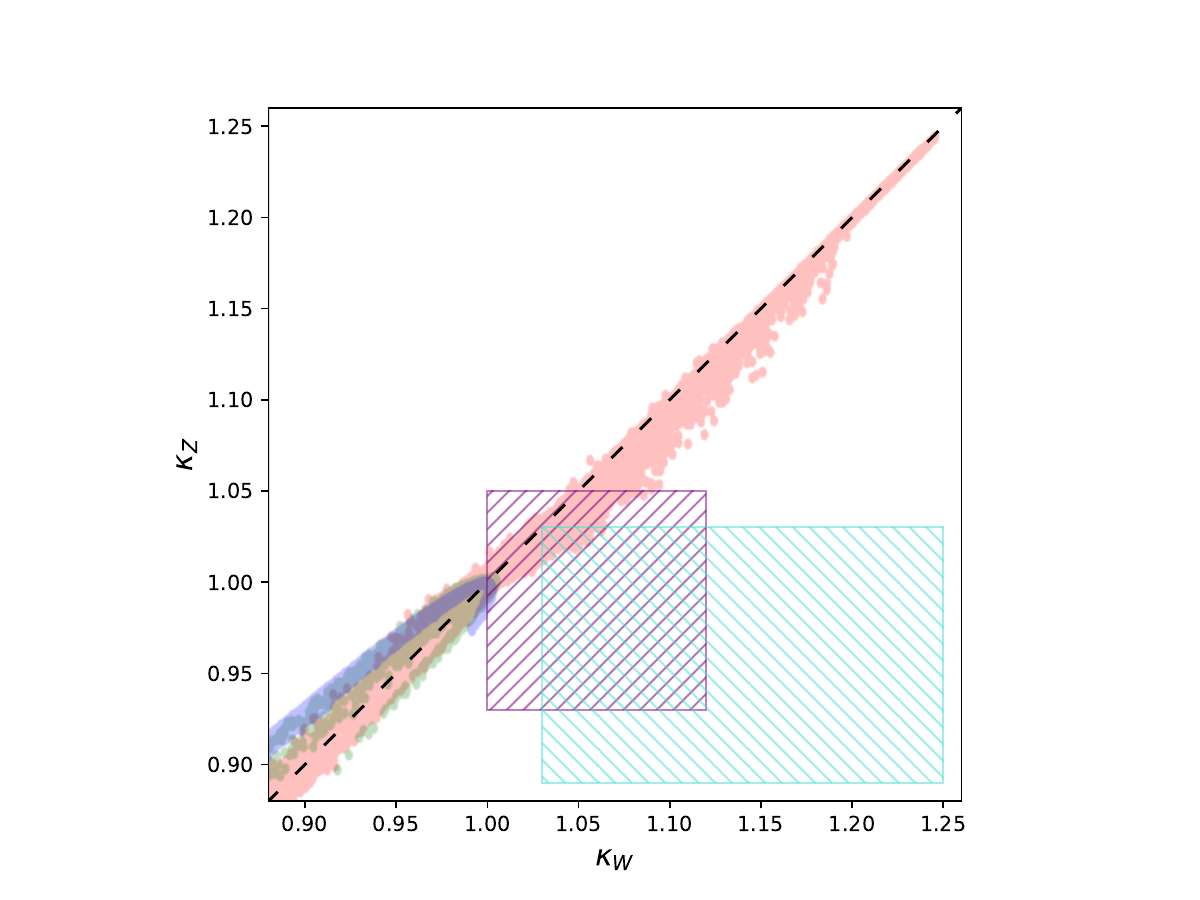}
\caption{Predicted values by the extended GM model with misaligned triplet mass terms in the $\kappa_W$-$\kappa_Z$ plane.  The red, green and blue points show the case with $v_\chi\in[5,50]$~GeV, $v_\chi\in[0.5,5]$~GeV and $v_\chi\in[0,0.5]$~GeV, respectively.  The dashed line indicates the custodial symmetric limit. The purple and cyan boxes indicate the 1$\sigma$ bounds given by the ATLAS~\cite{ATLAS:2021vrm} and CMS~\cite{CMS:2020gsy} measurements, respectively.
}
\label{ghZZ_vs_ghWW} 
\end{figure}

The predicted values of $\kappa_W$ and $\kappa_Z$ from the scan are shown in Fig.~\ref{ghZZ_vs_ghWW}, 
where the red, green and blue points are allowed for the case with $v_\chi\in[5,50]$, [0.5,5], and [0,0.5]~GeV, respectively. 
We see that $\kappa_W$ tends to be larger than $\kappa_Z$ for $v_\chi > 5$~GeV, which corresponds to the region slightly below the custodial symmetric limit indicated by the dashed line.  
This tendency is favored by the measurements of ATLAS~\cite{ATLAS:2021vrm} and CMS~\cite{CMS:2020gsy}\footnote{According to Ref.~\cite{CMS:2020gsy}, a negative value of $\kappa_W^{}$ is preferred by the combination of various production and decay channels of the Higgs boson. 
We here refer only to the magnitude of the $\kappa_W^{}$ value. }: 
\begin{align}
&\kappa_Z = 0.99\pm0.06, ~\kappa_W = 1.06\pm0.06 ~(\text{ATLAS}) , \\
&\kappa_Z = 0.96\pm0.07, ~|\kappa_W| = 1.11^{+0.14}_{-0.08} ~(\text{CMS}) .
\end{align}
On the other hand, for $v_\chi < 5$ GeV, most of the points appear at $\kappa_V < 1$ with $\kappa_Z^{} \gtrsim \kappa_W^{}$. 
Therefore, our scenario is well distinguished from the minimal triplet model if $\kappa_V$ is determined to be larger than unity and/or $\kappa_W^{} \gtrsim \kappa_Z^{}$ is confirmed by future experiments.

\begin{figure}[htbp]
\centering
\includegraphics[width=0.7\textwidth]{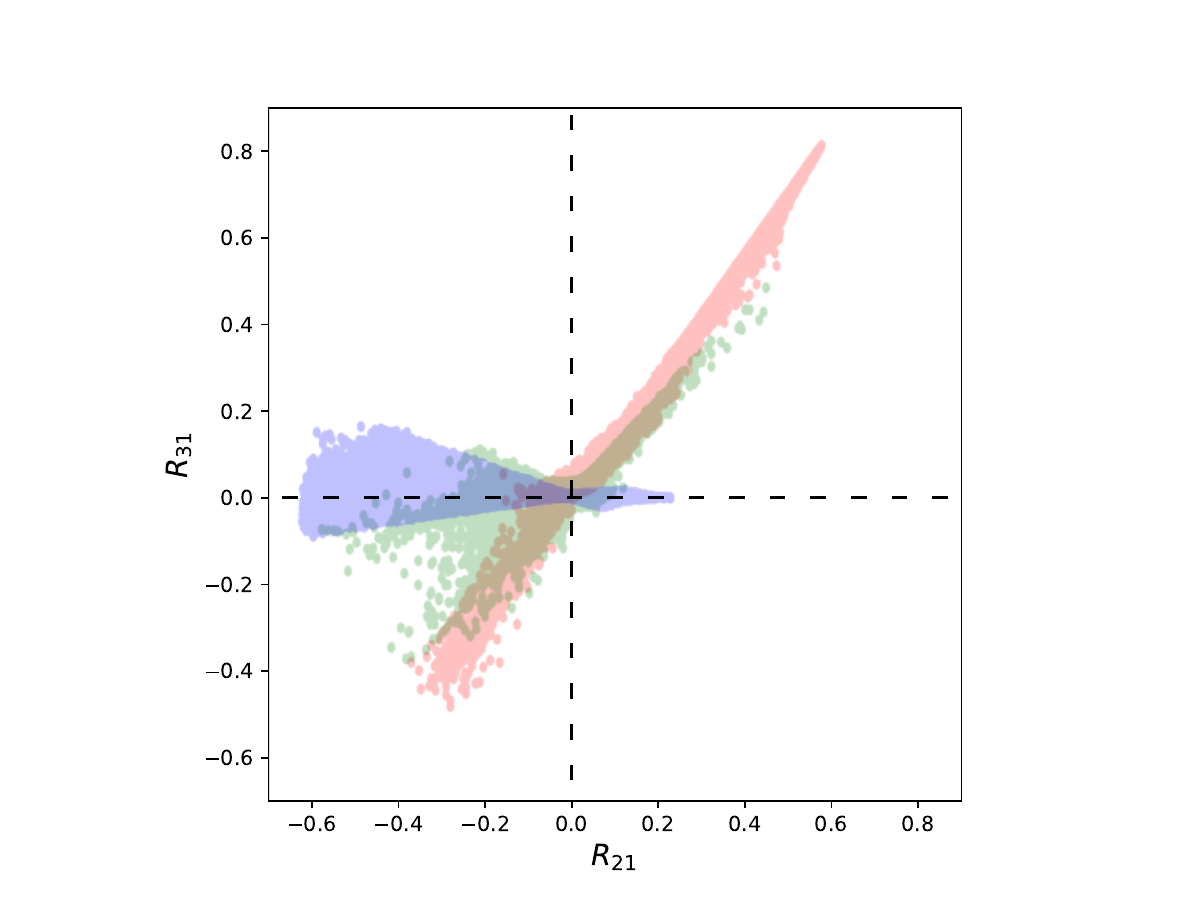}
\\
\caption{\label{R_all} 
Predicted values by the extended GM model with misaligned triplet mass terms in the $R_{21}$-$R_{31}$ plane. 
The red, green and blue points show the case with $v_\chi\in[5,50]$~GeV, $v_\chi\in[0.5,5]$~GeV and $v_\chi\in[0,0.5]$~GeV, respectively. 
}
\end{figure}

In order to clarify this behavior, we concentrate on the scenario with $v_\chi=0$ and $v_\xi\ll v_\Phi$, as $v_\xi \sim 4.6 $~GeV is required by CDF II measurement. 
In this case, we obtain
\begin{align}
\kappa_W^{} - \kappa_Z^{} = 4R_{21} \frac{v_\xi}{v}
\end{align}
from Eqs.~(\ref{kW}) and (\ref{kZ}), so that the sign of $R_{21}$ determines the relative magnitudes of $\kappa_W^{}$ and $\kappa_Z^{}$. 
The squared mass matrix for the $CP$-even Higgs bosons is given in the basis of $(h_\phi,h_\xi,h_\chi)$ as 
\begin{equation}
	M^2 = \begin{pmatrix}
		8\lambda_1v_\Phi^2 & (4\lambda_4-\lambda_5)v_\Phi v_\xi & 0 \\
		(4\lambda_4-\lambda_5)v_\Phi v_\xi & \frac{\lambda_5}{2}v_\Phi^2 + 8(\lambda_2+\lambda_3)v_\xi^2 & \frac{\lambda_5}{\sqrt{2}}v_\Phi^2 \\
		0 & \frac{\lambda_5}{\sqrt{2}}v_\Phi^2 & m_\chi^2 + \frac{4\lambda_4+\lambda_5}{2}v_\Phi^2 + 6\mu_2v_\xi+4\lambda_2v_\xi^2
	\end{pmatrix} . \label{eq:cp-even}
\end{equation}
By keeping terms to the leading order in $v_\xi^2 / v^2$, we obtain the expression for the rotation matrix $R$ as 
\begin{align}
R_{11} &= 1 + {\cal O}\left(\frac{v_\xi^2}{v^2}\right)
, \\
R_{21} &= -\left(\frac{\cos^2\theta}{m_{H_2}^2 - m_h^2}  + \frac{\sin^2\theta}{m_{H_3}^2 - m_h^2}\right) v_\Phi v_\xi(4\lambda_4-\lambda_5) + {\cal O}\left(\frac{v_\xi^2}{v^2}\right)
, \\
R_{31} &= \left(-\frac{\cos\theta\sin\theta}{m_{H_2}^2 - m_h^2}  + \frac{\cos\theta\sin\theta}{m_{H_3}^2 - m_h^2}\right) v_\Phi v_\xi(4\lambda_4-\lambda_5) + {\cal O}\left(\frac{v_\xi^2}{v^2}\right)
, 
\end{align}
where the mixing angle $\theta$ is given by
\begin{align}
\tan 2\theta = \frac{2M_{23}^2}{M_{22}^2 - M_{33}^2}
~. 
\end{align}
With the assumption that $m_h < m_{H_{2,3}}$, the sign of $R_{21}$ is completely determined by $4\lambda_4-\lambda_5$. This turns out to be highly constrained by one of the vacuum stability conditions:
\begin{equation}
	\lambda_4 >
\begin{cases}
	-\frac{1}{2}\lambda_5 - 2\sqrt{\lambda_1\left(\frac{1}{3}\lambda_3+\lambda_2\right)} \text{ for }\lambda_5\leq0 \text{ and }\lambda_3\geq 0 , \\
	-\omega_+(\zeta)\lambda_5 - 2\sqrt{\lambda_1\left(\zeta\lambda_3+\lambda_2\right)} \text{ for }\lambda_5\leq0 \text{ and }\lambda_3<0 , \\
	-\omega_-(\zeta)\lambda_5 - 2\sqrt{\lambda_1\left(\zeta\lambda_3+\lambda_2\right)} \text{ for }\lambda_5>0 ,
\end{cases}
\end{equation}
where the details of $\omega_{\pm}$ and $\zeta$ can be found in Ref.~\cite{Hartling:2014zca}. 
We show in Fig.~\ref{R_all} the scatter points in the $R_{21}$-$R_{31}$ plane for the three $v_\chi$ intervals.  It is found that in most of the cases $R_{21} < 0$ (corresponding to $4\lambda_4-\lambda_5 > 0$) at $v_\chi \simeq 0$, as indicated by the blue region in Fig.~\ref{R_all}.  Thus, $\kappa_Z^{} > \kappa_W^{}$ is favored for smaller values of $v_\chi$.  On the other hand, for $v_\chi>0.5$~GeV, $R_{21}$ and $R_{31}$ are mostly positively correlated.

We remark here that 
the decoupling limit, i.e., all the masses of the extra Higgs bosons become infinity and all the $H_1$ couplings with SM fields coincide with those of the SM values, 
can be realized by taking the limit $v_\chi \to 0$ and $v_\xi \to 0$ while keeping the ratio $r \equiv v_\chi/v_\xi $ to be finite.
In this limit, the mass matrix for the $CP$-even Higgs bosons takes the following form:
\begin{equation}
	M^2 = \begin{pmatrix}
		8\lambda_1v_\Phi^2 & 0 & 0 \\
		0 & r m_\chi^2 + \frac{4r\lambda_4+(1-r)\lambda_5}{2}v_\Phi^2 & \frac{\lambda_5}{\sqrt{2}}v_\Phi^2 \\
		0 & \frac{\lambda_5}{\sqrt{2}}v_\Phi^2 & m_\chi^2 + \frac{4\lambda_4+\lambda_5}{2}v_\Phi^2
	\end{pmatrix} .
\end{equation}
Thus, it is clear that only the $H_1$ state stays at the EW scale, while the other two states are decoupled in the limit of $m_\chi^2 \to \infty$ in which case $m_\xi^2$ also approaches infinity as required by the tadpole conditions. 
Similarly, the masses of the doubly-charged, singly-charged, and $CP$-odd states contain the $m_\chi^2$ term, so that they are also decoupled from the theory in the limit of $m_\chi^2 \to \infty$. 
This argument is consistent with that given in Ref.~\cite{Hartling:2014zca}.

\begin{figure}[htbp]
\centering
\includegraphics[width=0.7\textwidth]{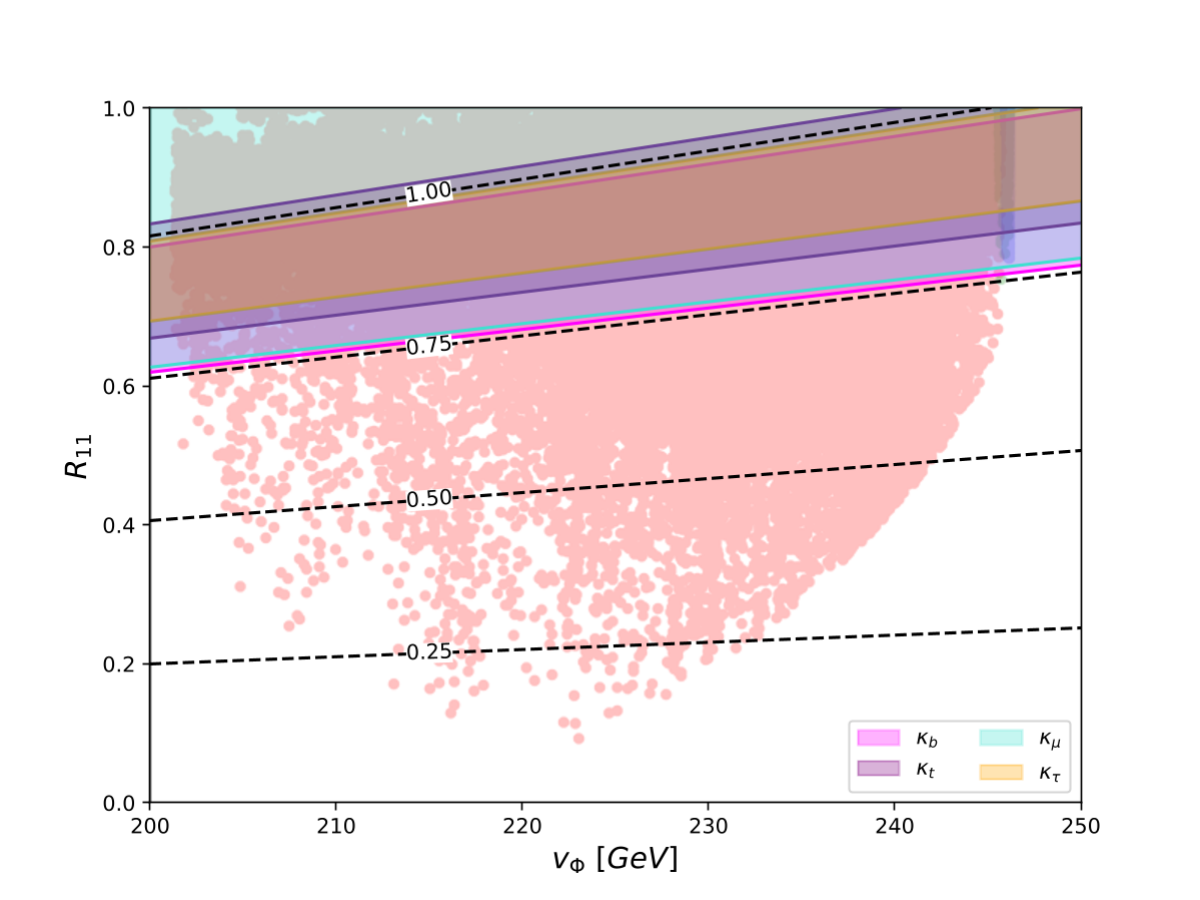}
\caption{\label{chphir_vs_vphi} Predicted values by the extended GM model with misaligned triplet mass terms in the $v_\Phi$-$R_{11}$ plane. The red, green and blue points show the case with $v_\chi\in[5,50]$~GeV, $v_\chi\in[0.5,5]$~GeV and $v_\chi\in[0,0.5]$~GeV, respectively. The contours for different values of $\kappa_f$ are denoted by the dashed curves. The different colored regions indicate the 1$\sigma$ bounds of $\kappa_{b,t,\mu,\tau}$ given by the ATLAS measurements~\cite{ATLAS:2021vrm}.}
\end{figure}

In Fig.~\ref{chphir_vs_vphi}, we further show the predicted values of $\kappa_f$, given by
\begin{equation}
	\kappa_f = R_{11} \frac{v}{v_\Phi} ,
\end{equation}
from the three scan ranges of $v_\chi$.
It is observed that the red region spans a significantly wider range from $\sim 0.25$ to $\gtrsim 1.00$, while the green and blue regions are restricted to the upper right corner since $v_\Phi\sim v$ and the mixing of $h$ with the other fields is mostly suppressed in these two cases.
Also shown in the plot are the contours of various $\kappa_f$'s in their 1$\sigma$ ranges measured by the ATLAS Collaboration~\cite{ATLAS:2021vrm}:
\begin{equation}
	\kappa_b=0.87\pm0.11, ~
	\kappa_t=0.92\pm0.10, ~
	\kappa_\mu=1.07^{+0.25}_{-0.30}, ~
	\kappa_\tau=0.92\pm0.07 .
\end{equation}
In both Figs.~\ref{ghZZ_vs_ghWW} and \ref{chphir_vs_vphi}, we point out the novel feature that, unlike the SM extended with scalar singlets and/or doublets, the model here can accommodate the possibilities that $\kappa_{W,Z}$ and/or $\kappa_f$ is greater than unity. On the other hand, if $v_\chi\to0$, then $\kappa_{Z,W,f}$ can barely exceed unity, and most of the time $\kappa_Z>\kappa_W$, as discussed before.

We note in passing that compared to the above-mentioned $\kappa$'s, the values of $\kappa_{\gamma}$ and $\kappa_{Z\gamma}$ depend on more unspecified parameters in the model and, therefore, present larger uncertainties, albeit there is a certain correlation between the quantities.  If we further take $\kappa_\gamma = 1.06 \pm 0.05$~\cite{ATLAS:2021vrm} as an input, our model predicts that $\kappa_{Z\gamma}$ would fall in the range of $(0.95,1.15)$, in comparison with the current ATLAS measurement of $1.43^{+0.31}_{-0.38}$~\cite{ATLAS:2021vrm}.

\begin{figure}[htbp]
\centering
\includegraphics[width=0.54\textwidth]{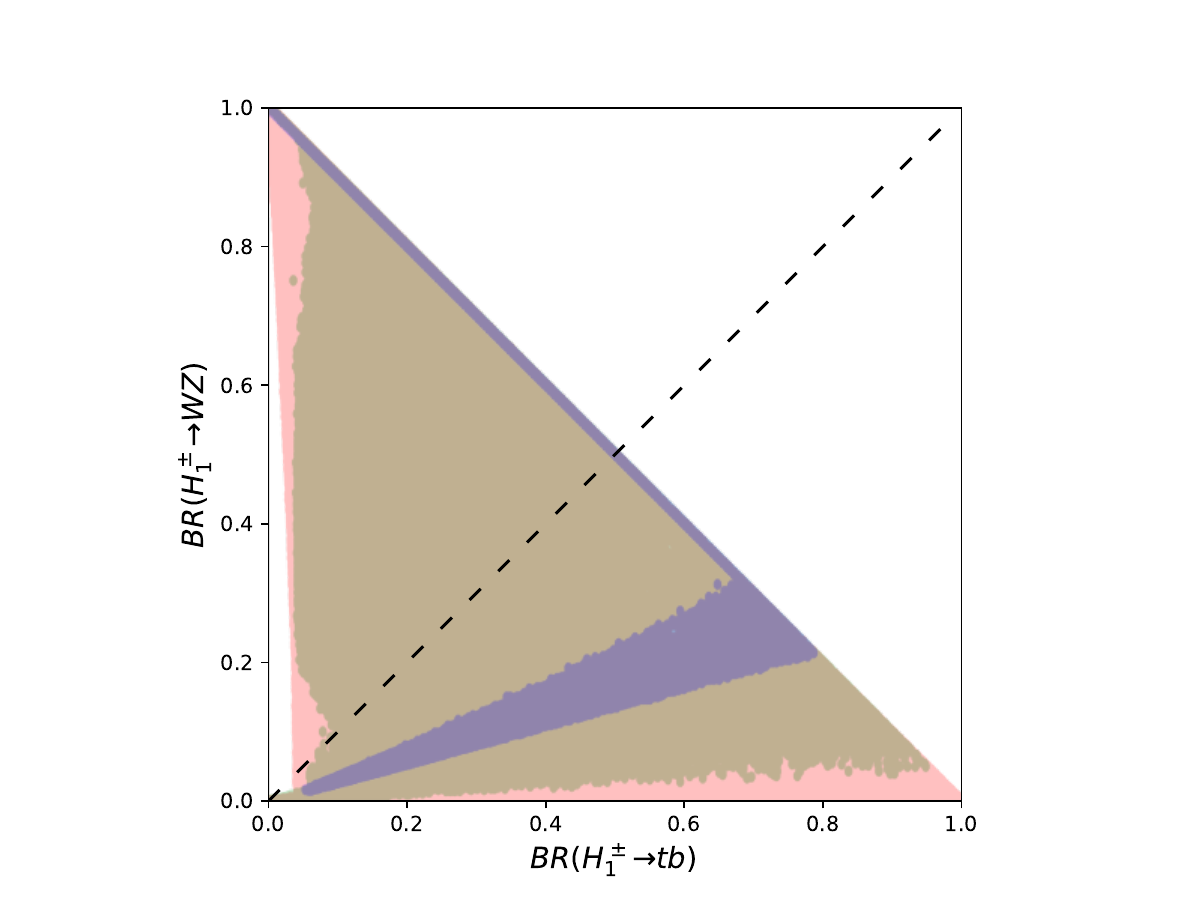}
\hspace{-1.8cm}
\includegraphics[width=0.54\textwidth]{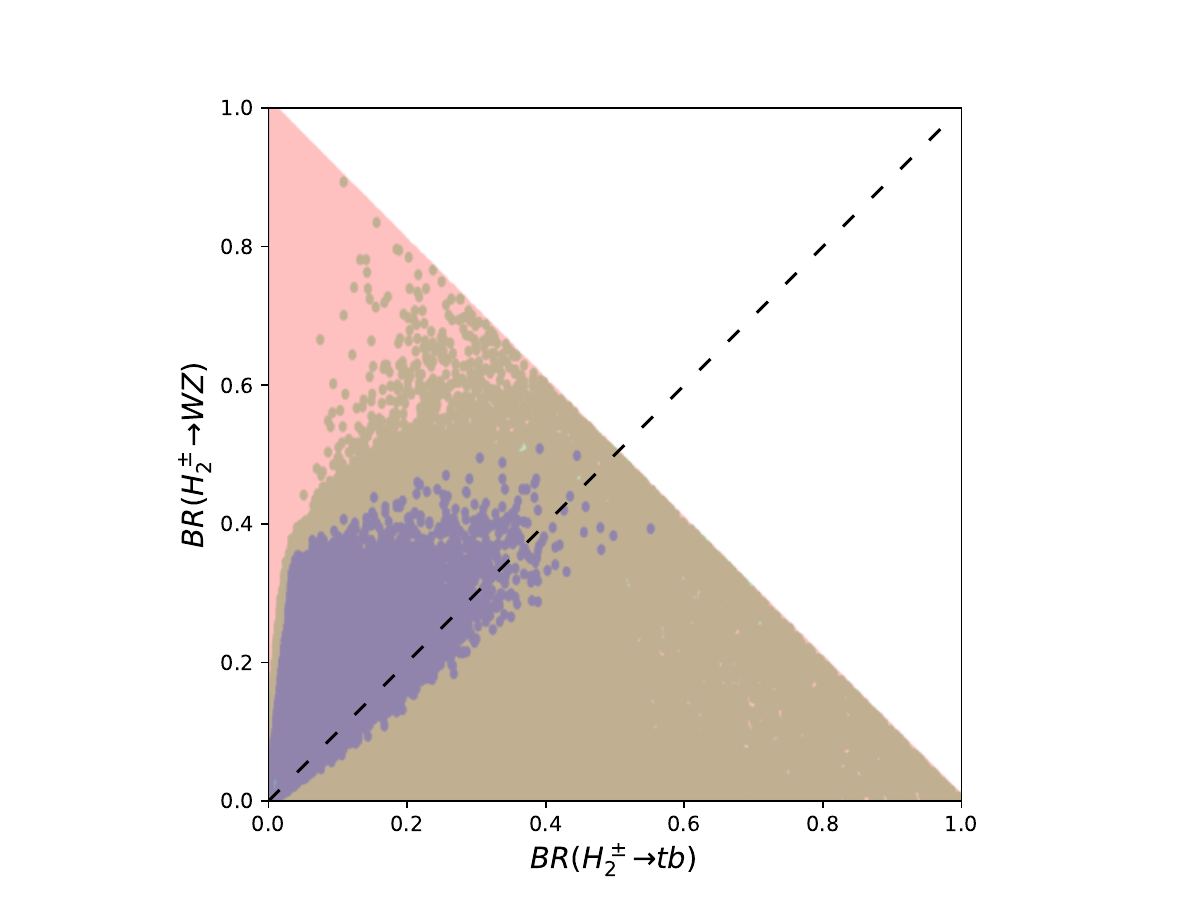}
\caption{\label{chargedHBRs} 
Predicted branching ratios of the $tb$ and $WZ$ decay channels for the $H_1^\pm$ boson (left plot) and the $H_2^\pm$ boson (right plot).  The red, green and blue points show the case with $v_\chi\in[5,50]$~GeV, $v_\chi\in[0.5,5]$~GeV and $v_\chi\in[0,0.5]$~GeV, respectively.}
\end{figure}

In the original GM model, the Higgs triplet $H_3^{0,\pm}$ and the Higgs quintet $H_5^{0,\pm,\pm\pm}$ are gaugephobic and fermiophobic, respectively.  In the extended GM model where the custodial symmetry is explicitly broken, the singly charged states can mix with each other.  As a consequence, both singly charged Higgs bosons can decay into a pair of gauge bosons and a pair of fermions.  (If $CP$-violating terms are further introduced, the neutral states will mix as well.)  In Fig.~\ref{chargedHBRs}, we present the branching ratios of the $tb$ and $WZ$ decay channels for $H_1^\pm$ (left plot) and $H_2^\pm$ (right plot).  They show that these two decay modes can potentially be summed up to be the dominant ones of the two charged Higgs bosons, as indicated by the points distributed along the diagonal.  Again, due to other unspecified parameters, the other decay modes such as $H^{\pm\pm}W^{\mp}$, $H^\pm Z$, and $H^{\pm\pm}H^{\mp}$ could be more dominant as well.

\section{Conclusions \label{sec:Conclusions}}

We reiterate that in order to accommodate the CDF II $W$ boson mass anomaly in a GM-based model, one has to extend by introducing custodial symmetry-breaking terms into the scalar potential.  At the same time, one is able to consider radiative corrections in a self-consistent framework.  In this case, there is the freedom to choose the counterterm for the EW $\rho$ parameter in such a way that the predicted value of $\rho$ matches with the measured one.  Based upon the CDF II result, the VEVs of the two triplet fields are found to satisfy $v_\xi^2 - v_\chi^2 \simeq$ $21.34\pm 2.52$~GeV$^2$.  This, in turn, implies $\kappa_W > \kappa_Z$ in most of the theoretical parameter space of the model.

\vspace*{4mm}

\begin{acknowledgments}
We thank H.~Beauchesne and C.~T.~Hsu for useful discussions.  T.-K. C. and C.-W. C. were supported in part by Grants No.~MOST-108-2112-M-002-005-MY3 and No. MOST-111-2112-M-002-018-MY3.  K. Y. was supported in part by the Grant-in-Aid for Early-Career Scientists, No.~19K14714.
\end{acknowledgments}

\bibliographystyle{apsrev}



\end{document}